\documentclass[journal]{IEEEtran}

\usepackage[cmex10]{amsmath}
\usepackage{amsfonts,amssymb,bm}

\usepackage{graphicx,color,psfrag}		
\usepackage{cite}			
\usepackage[capitalise]{cleveref}
\crefname{equation}{}{} 
\crefname{section}{Sec.}{Sec.}

\usepackage{epstopdf}	
\usepackage{enumitem} 

\graphicspath{{.},{figures/}}

\newcommand{\costheur}{J_\text{heur}}
\newcommand{\costprin}{J_\text{norm}}
\newcommand{\cost}{J}
\newcommand{\costspec}{L} 
\newcommand{\obs}{\hat{\cost}}
\newcommand{\param}{\theta}
\newcommand{\minparam}{\theta_\text{min}}
\newcommand{\uoparam}{\theta_{\star}}
\newcommand{\acqui}{\alpha}
\newcommand{\T}{\mathrm{T}}

\newcommand{\qmin}{\hat{p}_\text{min}}

\newcommand{\qminpred}{q_\text{min}(\uoparam)}

\newcommand{\pmin}{p_\text{min}}
\newcommand{\pmint}{p_\text{min}(\param)}

\newcommand{\argmax}{\operatornamewithlimits{argmax}}
\newcommand{\argmin}{\operatornamewithlimits{argmin}}
\newcommand{\dom}{\mathcal{D}}

\newcommand{\Hp}[1]{H_{#1}}
\newcommand{\ie}{i\/.\/e\/.,\/~}
\newcommand{\eg}{e\/.\/g\/.,\/~}

\newcommand{\fig}{Fig\/.\/~}

\newcommand{\sect}{Sec\/.\/~}

\newcommand{\paramdom}{\mathcal{D}}

\newcommand{\transp}{\text{T}}
\newcommand{\R}{\mathbb{R}}

\newcommand{\nx}{n}    
\newcommand{\nin}{{n_u}}    
\newcommand{\nout}{{n_y}}    
\newcommand{\ndist}{{n_d}}    
\newcommand{\nref}{{n_r}}

\DeclareMathOperator{\obspoles}{\mathcal P_{obs}}
\DeclareMathOperator{\ctrpoles}{\mathcal P_{ctr}}
\DeclareMathOperator{\ctrmatrix}{\mathcal K_{ctr}}

\makeatletter
\def\ps@IEEEtitlepagestyle{%
  \def\@oddfoot{\mycopyrightnotice}%
  \def\@evenfoot{}%
}
\def\mycopyrightnotice{%
  \begin{minipage}{\textwidth}%
  \footnotesize%
  \textbf{Accepted final version. }%
  To appear in \textit{IEEE Transactions on Control Systems Technology}.\\
	\copyright~2018 IEEE. Personal use of this material is permitted. Permission from IEEE must be obtained for all other uses, in any current or future media, including reprinting/republishing this material for advertising or promotional purposes, creating new collective works, for resale or redistribution to servers or lists, or reuse of any copyrighted component of this work in other works.
  \end{minipage}%
  \gdef\mycopyrightnotice{}
}

\begin{document}

\title{Data-efficient Auto-tuning with Bayesian Optimization: An Industrial Control Study}

\author{Matthias~Neumann-Brosig,
  Alonso~Marco,~\IEEEmembership{Student Member,~IEEE,}
  Dieter~Schwarzmann,
  and~Sebastian~Trimpe,~\IEEEmembership{Member,~IEEE}
\thanks{This work was supported in part by IAV GmbH, the Max Planck Society, and the Cyber Valley initiative.}
\thanks{M.~Neumann-Brosig and D.~Schwarzmann are with the TM Powertrain Mechatronics department at IAV GmbH, 38518 Gifhorn, Germany (e-mail: matthias.neumann-brosig@iav.de, dieter.schwarzmann@gmail.com).}
\thanks{A.~Marco and S.~Trimpe are with the Intelligent Control Systems group at the Max Planck
Institute for Intelligent Systems, 70569 Stuttgart, Germany (e-mail: alonso.marco@tuebingen.mpg.de, trimpe@is.mpg.de).}
}

\markboth{Authors' version}
{Authors' version}

\maketitle

\begin{abstract}
Bayesian optimization is proposed for automatic learning of optimal controller parameters from experimental data.  A probabilistic description (a Gaussian process) is used to model the unknown function from controller parameters to a user-defined cost.  The probabilistic model is updated with data, 
which is obtained by testing a set of parameters on the physical system and evaluating the cost.  In order to learn fast, the Bayesian optimization algorithm selects the next parameters to evaluate in a systematic way, for example, by maximizing information gain about the optimum.
The algorithm thus iteratively finds the globally optimal parameters with only few experiments. Taking throttle valve control as a representative industrial control example, the proposed auto-tuning method is shown to outperform manual calibration: it consistently achieves better performance with a 
low number of experiments.
The proposed auto-tuning framework is flexible and can handle different control structures and objectives.

\end{abstract}


\IEEEpeerreviewmaketitle

\section{Introduction}
\label{sec:intro}
The need for tuning of controller parameters is ubiquitous in industry.  Virtually every controller involves a number of parameters whose proper choice is critical for performance.  However, 
the process of finding good parameters is often expensive, for example, 
consuming significant time from the operator or engineer.  Because tuning controllers manually is tedious and involved, controllers often operate at low performance or not at all (see \cite{Je06} and references therein).
The strong need for automatic controller tuning algorithms is thus apparent.

An ideal auto-tuner combines the following characteristics: it is 
\emph{versatile} 
(\ie applies to various control structures), 
\emph{globally optimal} (finds the best controller), 
and \emph{data efficient} (requires little experimental time on the plant).
Most existing methods meet some, but not all of these objectives.  
Classical automatic tuning methods \cite{AsHaHaHo93,Ha15} target simple, typically single-loop controller structures such as PID control. 
While optimization-based techniques such as gradient-based approaches (\eg \cite{Hj02,KiKr06} and references therein)
or evolutionary search \cite{FlPu02} are applicable to general controller tuning problems in principle, they yield only local optimality in the former case and often require an impracticably large number of experiments.

\begin{figure}[tb]
\centering
\includegraphics[width=0.8\columnwidth]{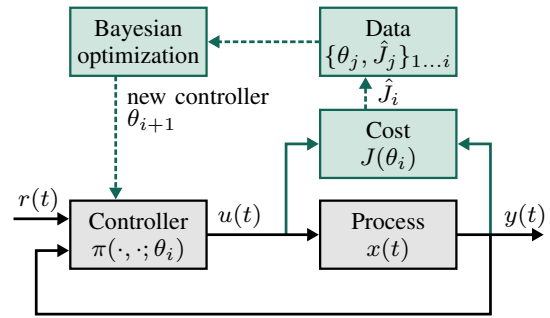}
\caption{Proposed framework using Bayesian optimization (BO) for automatic controller tuning.  The controller parameterization $\theta_i$ is evaluated in closed-loop operation in terms of a cost functional $J$.  Based on all previous experiments  
$\{\theta_j, \hat{J}_j\}_{j=1,\dots,i}$, 
BO suggests the next controller $\theta_{i+1}$ aiming at finding the global optimum with only few iterations.  }
\label{fig:TuningFramework}
\end{figure}

In this article, we propose Bayesian optimization (BO) for automatic controller tuning 
(see \fig \ref{fig:TuningFramework}) and develop a framework for automatic learning of controller parameters that combines the above desiderata. 
With this framework, an optimal controller is obtained from only few experiments and without the need for a process model.
 
BO employs a probabilistic description (typically, a Gaussian process \cite{Rasmussen2006Gaussian}) of the underlying unknown objective function, \ie the map of control parameters to the user-specified control objective.
The Bayesian treatment allows for a principled combination of prior problem knowledge with experimental data.  The probabilistic description of the objective function captures, in each iteration, the current information about the tuning problem gathered from previous evaluations and prior knowledge.  This information is used to select the next experiment in a systematic way, for example, to maximize the information gain from each experiment in order to find the optimal controller with few evaluations. 
Even though some key ideas date back to the 50s and 60s \cite{krige1951statistical,kushner1964new}, BO has only recently gained a lot of popularity in the machine learning community 
empowered by today's computational power and novel efficient algorithms (see \cite{shahriari2016taking} for a recent overview).

We demonstrate the capability of BO for automatic and data-efficient tuning on a typical industrial controller.  We consider control of a throttle valve using active disturbance rejection control (ADRC) \cite{HUANG2014963} as the control structure.  Throttle valve control is a relevant problem in automotive industry since good controller parameters are crucial for performance and safety, and those need to be tuned for each application.
ADRC is a favorable controller structure in practice, because it is often intuitive and straightforward to tune for a human.  Nonetheless, we show herein that BO outperforms human controller tuning: BO achieves on par or better performance than hand tuning with only 
around ten experimental evaluations.

\subsubsection*{Contributions}
In detail, this article makes the following main contributions:
\begin{enumerate}
\item Proposal of Bayesian optimization for automatic tuning of controller parameters;
\item Development of a controller learning framework, which combines BO with ADRC and thus achieves an effective balance between exploiting prior structural knowledge (esp.\ dynamic system order) and learning from data;
\item Experimental demonstration of superior learning results (compared to hand tuning) with an extremely low number of samples; and
\item Evaluation of the proposed learning approach under different types of control objectives (\eg response speed, robustness).
\end{enumerate}
In view of $1)$ and $2)$: Even though we apply two different BO algorithms and cost functionals in our experiments to demonstrate the adequacy of the proposed tuning method, it is not our goal to systematically compare different BO algorithms against each other. Rather, we deliberately choose dissimilar algorithms (cf. \cref{subsec::EI} and \cref{subsec::ES}) and functionals (\cref{subsec::choiceofffunctionals}) to empirically strengthen the claims in $1)$ and $2)$. The comparison of different BO algorithms is an active research topic (\eg \cite{calandra2016bayesian,VoTrMaFiPa18}) and beyond the scope of this article.

\subsubsection*{Related work}
First algorithms for automatic controller tuning appeared
in the 70s and 80s with the advent and success of computer-based control in industry \cite{Ha15}.  Classical tuning methods mainly focus on step-response analysis or relay tuning; see \cite{AsHaHaHo93,Ha15} for overviews. 
While these methods aim at simple computational schemes, today's computer technology allows for considering powerful machine learning methods for controller tuning.
BO
falls into this category.

The problem of controller tuning is intimately connected to that of reinforcement learning (RL) \cite{ScAt10,KoBaPe13}.  In the RL context, the problem herein can be characterized as episodic policy search with a continuous state-action space and a known reward (or cost) function.

In both automatic tuning and RL, one distinguishes \emph{direct} (model-free) and \emph{indirect} (model-based) methods \cite{AsHaHaHo93,ScAt10}.  While in direct methods, controller parameters are directly adjusted from closed-loop data, indirect methods first build a dynamical model from data, which then serves as the basis for obtaining controller parameters.  While model-based methods can be beneficial, \eg in terms of data-efficiency, they strongly rely on the learned model accurately capturing the true dynamics, which by itself is a challenging problem \cite{doerr_corl_2017}.
The method herein is direct, thus sidestepping the potential difficulties of model learning.  Notwithstanding, we do include relevant dynamics parameters (poles of an approximate linear model) in the parametric tuning procedure, which facilitates the learning process. Recent methods on indirect tuning \cite{DeRa11,DeFoRa15,DoNgMaScTr17,doerr_corl_2017} leverage the same probabilistic learning framework (GPs) as the one used herein.
Whereas these methods train a GP dynamics model from data, we use a GP to capture the underlying objective function.

BO for controller learning has recently also been suggested in \cite{calandra2016bayesian,BeScKr16,MaHeBoScTr16}, which include successful demonstrations  
in laboratory experiments.
A discrete event controller is optimized for a walking robot in \cite{calandra2016bayesian}, and state-feedback controllers are tuned in \cite{BeScKr16} for a quadrotor and in \cite{MaHeBoScTr16} for a humanoid robot balancing a pole.
Herein, we present results of applying BO for a typical control problem in the automotive industry (throttle valve control) and consider two types of control objectives, different from those in \cite{calandra2016bayesian,BeScKr16,MaHeBoScTr16}.  The proposed controller learning framework, which combines BO with ADRC, is different from the controllers in the mentioned references.

An alternative for direct controller tuning 
is \emph{virtual reference feedback tuning} (VRFT), which aims at obtaining controller parameters from just one experiment.  VRFT was originally developed for linear systems \cite{CaLeSa02} with extensions to nonlinear single-input-single-output systems in \cite{CaSa06}.  This approach is based on specifying the control objective by means of a reference model. In contrast, the approach herein can handle general control objectives, and the consideration of multi-input-multi-output systems makes no difference.
Other alternatives for direct tuning include gradient-based approaches \cite{Hj02,KiKr06}, which rely on differentiability assumptions not needed here.  Unlike most tuning methods from the field of adaptive control, BO tuning does also not require a convex parametrization.  
Evolutionary search \cite{FlPu02} does not rely on differentiability and convexity assumptions either, but usually requires an impractically large number of evaluations.

Auto-tuning for throttle valve control is also considered in \cite{PaDeJaPe06,WaYuWaYa13,BiHgKoMaKn13}, for example.  In \cite{PaDeJaPe06}, a tuning routine dedicated to throttle valve control is developed, which is based on a special identification procedure of relevant process parameters.
Direct tuning of a fuzzy-PID controller via an evolutionary search algorithm is proposed in\cite{WaYuWaYa13}.  Bischoff \emph{et al.} \cite{BiHgKoMaKn13} consider auto-tuning of throttle valve control via stochastic learning techniques similar to those herein.  They apply the indirect/model-based  method \cite{DeRa11}, which employs GPs for learning dynamics models.

 \section{Learning Control Problem}
\label{sec:controlProblem}

We consider an uncertain nonlinear dynamic system 
\begin{align}
\dot{x}(t) &= f(x(t),u(t),d(t))  \label{eq:sys_state} \\[1ex]
y(t) &= h(x(t),d(t)) \label{eq:sys_meas}
\end{align}
where $x(t) \in \R^\nx$ is the state, $u(t) \in \R^\nin$ the control input, $y(t) \in \R^\nout$ the measured output, and $d(t) \in \R^\ndist$ represents some general disturbance (\eg process disturbance, sensor noise).  In the general learning control problem, $f$, $h$, $x$, and $d$ are unknown.  In the throttle valve example (\sect \ref{sec::throttle_valve_control}), we exploit \emph{some} structural knowledge about $f$ and $h$, while the complete dynamics remain unknown.

We seek an output-feedback controller to track a reference $r(t) \in \R^\nref$ with the system output $y(t)$; that is, 
\begin{equation}
u(t) = \pi(y(t), r(t); \theta)
\label{eq:control_general}
\end{equation}
where the controller $\pi$ 
can itself have internal states.
The controller is parametrized by $\param$ from some compact domain $\paramdom \subset \R^N$.  Given a controller structure $\pi$, we seek parameters $\minparam$ that optimize a given control objective such as minimizing overshoot,
a quadratic cost, or maximizing a robustness measure.
Generically, the control objective can be expressed as a cost functional $\costspec$ of input $u$, output $y$, and reference $r$, for example,
a quadratic cost
\begin{equation}
\costspec = \int_0^T \|y(t)-r(t)\|^2 + \|u(t)\|^2 \, dt
\label{eq:costspec}
\end{equation}
with the Euclidean norm $\| \! \cdot \! \|$.
Other examples are discussed in \sect \ref{sec::design_learning_experiments}.

For a given set of parameters $\theta$, the closed-loop response is determined through \eqref{eq:sys_state}, \eqref{eq:sys_meas}, and \eqref{eq:control_general}.  The definition of a cost functional $L$ thus fixes the function $J$ that maps parameters $\theta$ to cost values,
\begin{equation}
\cost: \mathcal D \rightarrow \mathbb R.
\label{eq:costFunction}
\end{equation}
We can then state tuning of the controller \eqref{eq:control_general} as an optimization problem: find parameters $\minparam \in \mathcal D$ such that
\begin{equation}
\forall \param \in \mathcal D: \cost(\minparam) \leq \cost(\param) .
\label{eq:learning_problem}
\end{equation}
Existence of a minimum $\minparam$ is assumed, although, in general, there may be more than one solution.

While the functional $\costspec$ is obviously known (it is specified by the designer), the function \eqref{eq:costFunction} is \emph{not}, because the dynamic system \eqref{eq:sys_state} and \eqref{eq:sys_meas} is unknown.  
We can sample $J$ by performing an experiment on the system and evaluating $\costspec$ from data $u$, $y$, and $r$ recorded over a suitable horizon.  
However, this sampling procedure is typically expensive (\eg involving monetary cost, operator time, or causing system wear and tear), which permits only few experiments and makes the optimization \eqref{eq:learning_problem} challenging.  What is more, the samples of $\cost$ are typically uncertain, for example, because of noisy sensor data.  In addition, the objective function is generally nonconvex and no gradients are readily available, even in the case where $\cost$ is differentiable.  
We thus need an optimization algorithm capable of dealing with these challenges.

 \section{Data-efficient Bayesian Optimization} 
\label{sec::BO}

Bayesian optimization (BO) denotes a class of algorithms for black-box global optimization problems in which data collection is expensive~\cite{kushner1964new} and thus, only few evaluations are possible. 
To deal with scarce data, BO (i) assumes a probabilistic prior about the objective function and (ii) chooses wisely the next combination of parameters to try on the system according to a pre-established \emph{acquisition function}.

In this section, we first introduce Gaussian processes, a probabilistic framework that allows for non-parametric regression (function approximation) on the unknown objective function. Second, we briefly introduce BO, and how it can be applied to the learning control problem presented in~\cref{sec:controlProblem}. Finally, we focus on two specific BO algorithms: \emph{Entropy search} \cite{HeSc12}, and \emph{expected improvement} \cite{jones1998efficient}. 

\subsection{Gaussian Processes (GPs)}
\label{ssec:GPs}
In this part, we give a brief introduction of GPs. Readers interested in a detailed explanation are referred to~\cite{Rasmussen2006Gaussian}.\tabularnewline

We use GPs to model the unknown cost function \cref{eq:costFunction} and to make probabilistic predictions about the cost function values at unobserved locations $\param$. A GP over $\cost$ is then defined as a collection of 
random variables
$\cost({\param})$, any finite number of which have a joint Gaussian distribution. GPs are considered a non-parametric probabilistic regression tool since no assumption is made on the parametric structure of $\cost$.

For a closed-loop experiment in \cref{eq:sys_state} and \cref{eq:sys_meas}, with controller parameters $\param_i$, we model noisy observations of \cref{eq:costFunction} as
\begin{equation}
\obs_i = \cost(\param_i) + \varepsilon_i
\label{eq:likelihood}
\end{equation}
with $\varepsilon_i \sim \mathcal{N}(0,\sigma_\text{n}^2)$. 

Prior knowledge about~$\cost$ can be included in the Gaussian process regression model through a prior mean function~$m\colon \mathcal{D} \rightarrow \mathbb{R}$, and a covariance function~$k\colon \mathcal{D} \times \mathcal{D} \rightarrow \mathbb{R}$. Whereas $m(\param)=\mathbb{E}\left[ J(\param) \right] $ represents the expected value of the function, $k(\param,\param')=\mathbb{C}\text{ov} \left[ J(\param),J(\param') \right]$ defines the covariance between two stochastic function values $\cost(\param)$ and $\cost(\param')$. The covariance function $k$, which is also called \emph{kernel}, encodes prior assumptions, \eg about smoothness and rate of change of $\cost$.

We can predict the performance of a closed-loop experiment with controller parameters $\param$, by computing its mean $\mu(\param)$ and variance $\sigma^2(\param)$ conditioned on a set of $N$ past observations $\{ \theta_i , \obs_i \}_{i=1}^N$, as
\begin{align}
	\mu(\param) &= m(\param) + k^\T(\param)  K^{-1} z
	\label{eq:gp_prediction_mean} \\
	\sigma^2(\param) &= k(\param,\param) - k^\T(\param) K^{-1} k(\param),
	\label{eq:gp_prediction_variance}
\end{align}
where 
$z$ and $k(\param)$ are column vectors with entries $[z]_{i}=\obs_i-m(\theta_i)$ and $[k(\param)]_{i}=k(\param,\param_i)$, ${K \in \mathbb{R}^{N \times N}}$ is the Gram matrix with entries ${[K]_{(i,j)} = k(\param_i, \param_j) + \delta_{ij} \sigma_\text{n}^2 }$, ${i,j\in\{1,\dots,N\}}$, and $\delta_{ij}$ is the Kronecker delta.

The covariance function $k$ depends on a set of parameters, e.g., lengthscales and prior signal variance, which can be chosen to determine the shape of the functions that $k$ encodes. These, combined with the variance of the evaluation noise $\sigma_\text{n}^2$, are typically called \emph{hyperparameters} of the GP. 
Alternative to specifying fixed hyperparameters, they can be estimated from data via marginal likelihood maximization \cite{Rasmussen2006Gaussian}. 
Also, hyperparameter treatment has been addressed in \cite{snoek2012practical} and \cite{gelbart2014bayesian}, where the acquisition function is marginalized over hyperparameters using Markov Chain Monte Carlo methods.

\begin{figure}[bt!]
\centering
\includegraphics[width=\columnwidth]{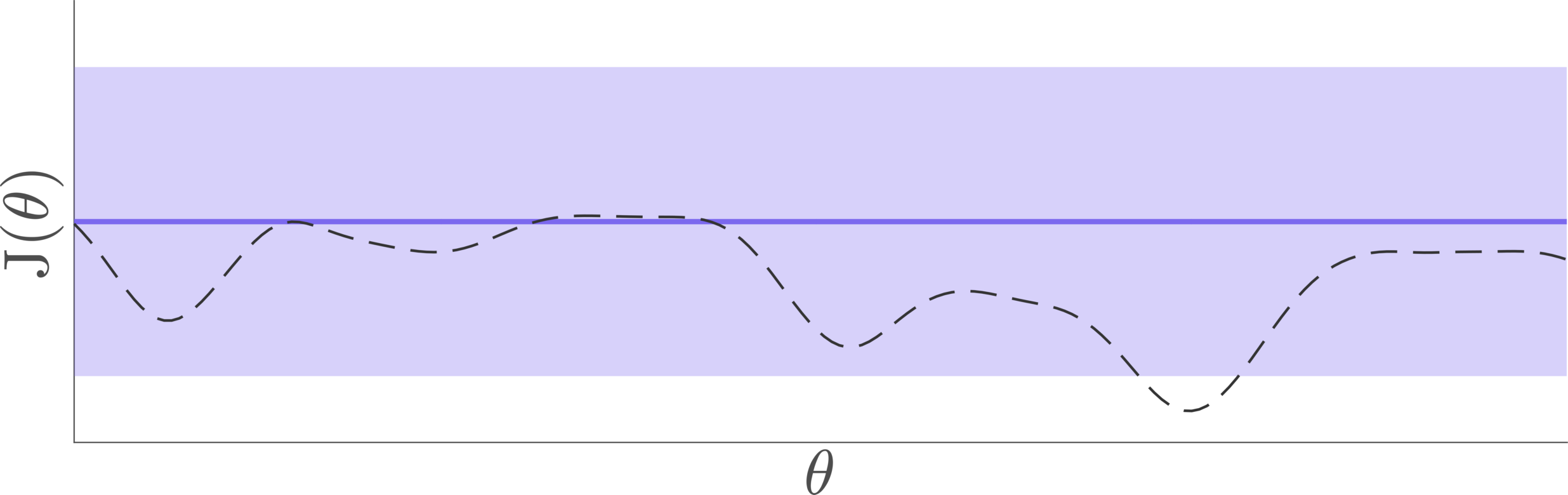}
\caption{GP prior before observing any data. The unknown function (dashed line) is to be learned from data. The prior mean function (solid line) is assumed to be zero here. The prior signal variance (colored surface) is represented with $\pm$ two standard deviations of the prior distribution. Thus, the unknown function is expected to be contained within the surface limits with a 95$\%$ confidence. }
\label{fig:GP_nodata}
\end{figure}

\begin{figure}[bt!]
\centering
\includegraphics[width=\columnwidth]{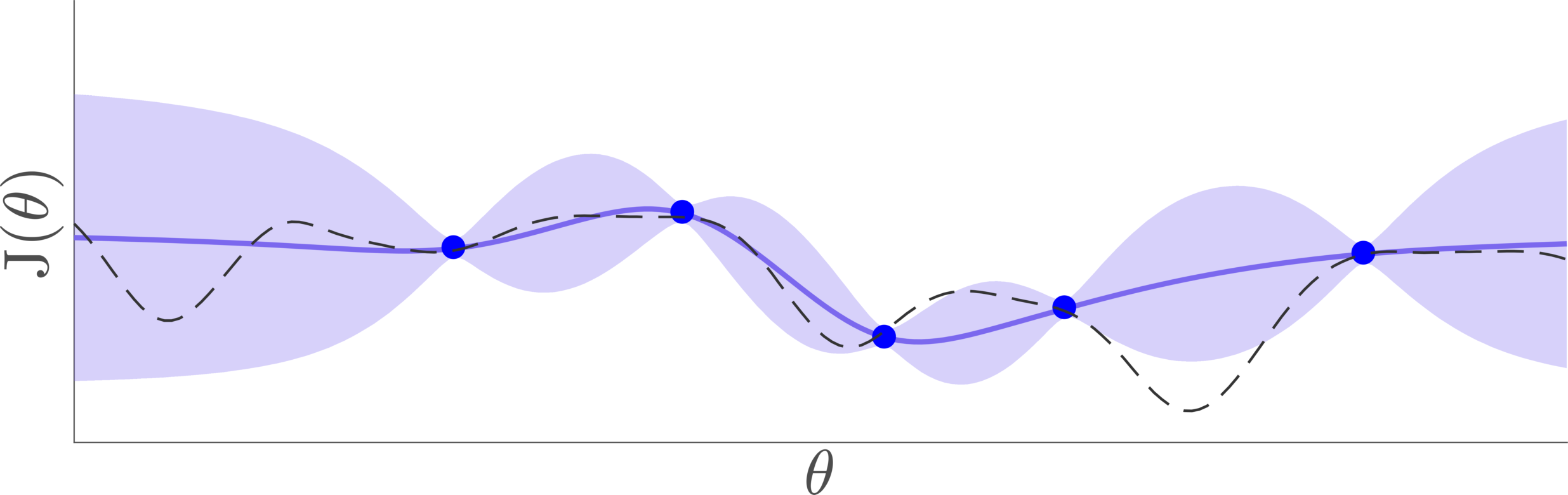}
\caption{GP posterior, after five noisy evaluations (dots). The posterior mean \cref{eq:gp_prediction_mean} is updated by conditioning on the observed data. The posterior variance \cref{eq:gp_prediction_variance} shrinks locally around the data points, representing reduced uncertainty. The remaining elements of this plot are described in~\cref{fig:GP_nodata}.}
\label{fig:GP_somedata}
\end{figure}

\cref{fig:GP_nodata} exemplifies a Gaussian process in a one dimensional regression problem, before observing any data. The prior mean function $m$ is assumed to be zero. The colored surface represents the prior signal variance as two standard deviations. After observing five data points, we can compute the Gaussian posterior,
as shown in~\cref{fig:GP_somedata}.

\subsection{Bayesian optimization (BO)}
\label{ssec:BO}
BO uses the GP model of the objective function to systematically select next function evaluations in order to find the global minimum efficiently. Next, we discuss the main steps taken by a Bayesian optimizer at each iteration.

\begin{figure}[bt!]
\centering
\includegraphics[width=\columnwidth]{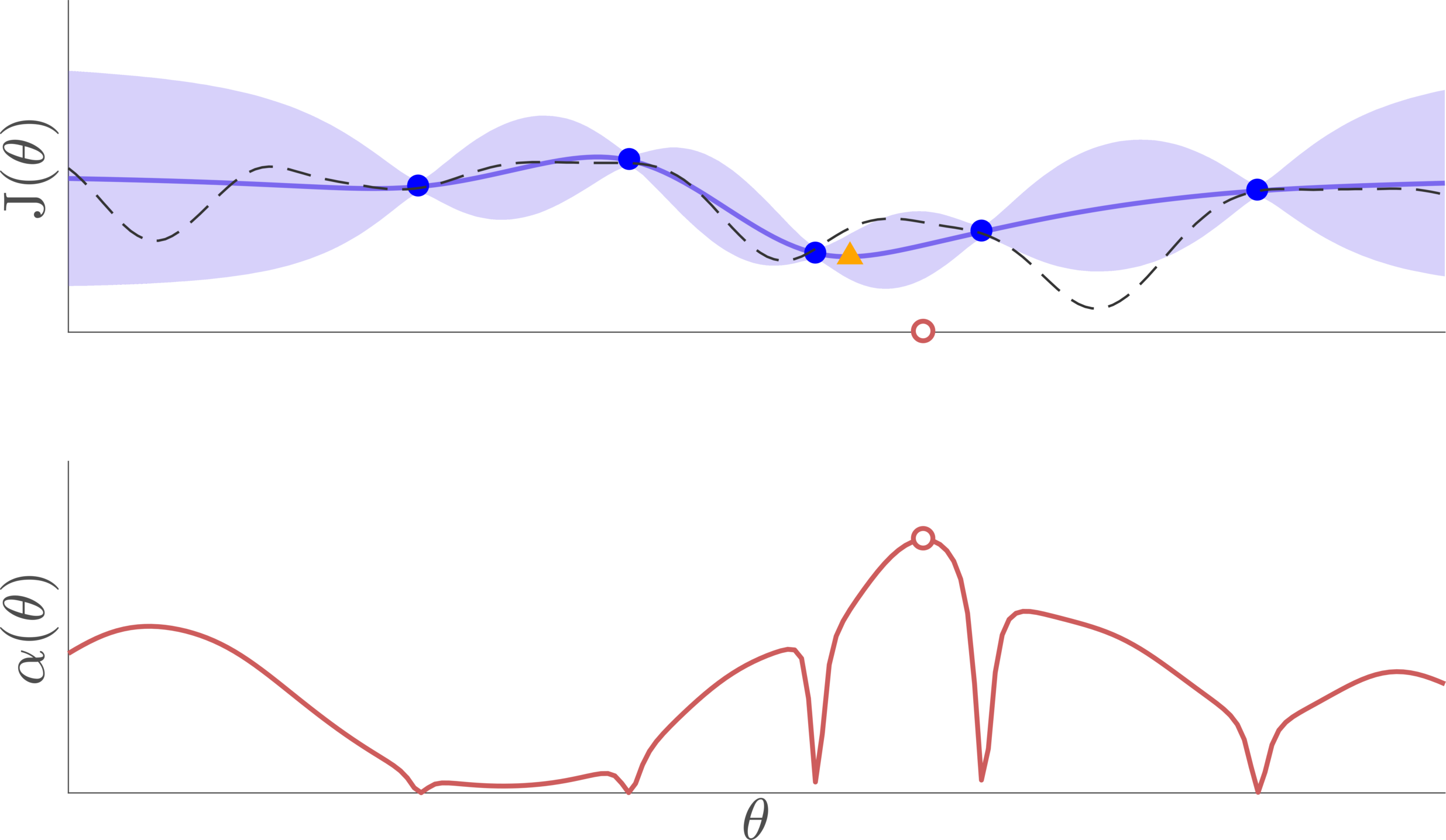}
\caption{Intermediate stage of a BO algorithm, before finding the true minimum. At the top, GP posterior conditioned on observed data. The current estimate of the global minimum (triangle) does not coincide with the true minimum yet. At the bottom, acquisition function $\acqui$ computed by the BO algorithm. The next parameters $\param_\text{next}$ are chosen at the maximum of $\acqui$ (hollow red dot).}
\label{fig:BO}
\end{figure}

The acquisition function $\acqui\colon \mathcal{D} \rightarrow \mathbb{R}$ takes into account the posterior after observations (\cref{fig:GP_somedata}) to suggest a location to acquire a new data point.
That is, the next parameters $\param_\text{next}$ are selected by maximizing the acquisition function as
\begin{equation}
\param_\text{next} = \argmax_{\param \in \dom} \acqui(\param).
\label{eq:acqui_max}
\end{equation}
While querying the objective function $J$ requires an expensive physical experiment, querying $\acqui$ involves executing an algorithm, usually computationally cheap. Thus, the problem $\cref{eq:acqui_max}$ can be solved using gradient-free or gradient-based optimization methods that require many function queries. \cref{fig:BO} (bottom) shows the acquisition function computed using the posterior GP model conditioned on the observed data.
The next parameters $\param_\text{next}$, selected at the maximum of $\acqui$, are used to perform a new experiment. Its outcome is then used to update the GP model, and to recompute $\acqui$ in the next iteration. This procedure is repeated until a suitable stopping criterion is satisfied. In our experiments, we terminate the search after a fixed number of evaluations.

What differentiates one BO algorithm from the other is the nature of the acquisition function. 
One of the most widely used is expected improvement (EI) \cite{jones1998efficient}, which selects $\param_\text{next}$ in order to improve over the solution found so far.
A more recent algorithm is entropy search (ES) \cite{HeSc12}, which chooses $\param_\text{next}$ as the most informative location about the global minimum. 
Other popular BO algorithms include probability of improvement (PI) \cite{lizotte2008} and upper confidence bound 
(GP-UCB) \cite{srinivas2009gaussian}, which are discussed and compared in \cite{shahriari2016taking}.
Whereas the tuning problem~\cref{eq:learning_problem} is agnostic to the selected BO method, its outcome (e.g., incurred number of experiments) may differ from 
one another.

Since the aforementioned methods are stochastic optimizers, the repeatability of their final outcome is affected by a number of aspects.
For instance, ES requires internal numerical approximations that are realized by sampling from distributions, which can influence the decisions made at each iteration, and thus, its outcome. In addition, the data itself is affected by noise,
which obviously impacts the GP posterior (and thus BO decision), as well as the outcome of hyperparameter optimization (both, if done in advance or online).

However, despite these sources of randomness, BO is able to quantify uncertainty and thus systematically address controller tuning in this stochastic setting.  We are interested in these methods as tools to improve hand tuning of controllers (with less experimental effort), which will face stochasticity in the data likewise, and may be considered even less repeatable when comparing across different engineers.

In this work, the controller parameters are learned using the ES and EI methods, which are  briefly introduced next. Readers interested in details are referred to \cite{HeSc12} and \cite{jones1998efficient} respectively.

\subsection{Expected improvement (EI)}\label{subsec::EI}
This criterion seeks next evaluations where we expect the objective function to improve most over the lowest cost function value $\eta$ collected so far. The acquisition function is defined as $\acqui(\param) = \mathbb{E}\left[ \max \left( 0,\eta - J(\param) \right) \right]$, which can be analytically solved \cite{jones1998efficient} as
\begin{equation}
\acqui(\param) = (\eta - \mu(\param))\Phi(z(\param))+\sigma(\param) \phi(z(\param))
\end{equation}
where $\Phi$ is the standard Gaussian cumulative density function, $\phi$ is the standard Gaussian probability density function, and $z(\param)=(\eta-\mu(\param))/\sigma(\param)$, with $\mu(\param)$ and $\sigma(\param)$ computed from~\cref{eq:gp_prediction_mean} and~\cref{eq:gp_prediction_variance} respectively.

\subsection{Entropy search (ES)}\label{subsec::ES}
Contrary to other methods (like EI, PI or GP-UCB), ES explicitly approximates a probability distribution about the location of the minimum at each iteration, and computes its entropy, i.e., information about the minimum. Then, the next location is selected where the expected information increment is maximal, i.e., where we expect to gain most information about the minimum. Finally, the global minimum is estimated for the current iteration.
These procedures can be divided in three steps, which we explain next on an abstract level; for details, we refer to the original paper~\cite{HeSc12}.

\subsubsection{Distribution about the location of the minimum}
ES models the knowledge about the location of the global minimum with a probability density $\pmin$ such that
\begin{equation}
\pmint = p(\param \in \argmin_{\tilde{\param} \in \dom} \cost(\tilde{\param})).
\label{eq:pmin}
\end{equation}
There are cases in which \cref{eq:pmin} cannot be strictly defined as a probability density. However, this is a minor issue in practice, as ES needs to approximate \cref{eq:pmin} as a probability distribution $\qmin$ to make it computationally tractable. We define $\qmin$ on an irregular grid over $\mathcal{D}$, which puts higher resolution in regions more likely to contain the minimum (see \cite{HeSc12} for details).

\subsubsection{Acquisition function}
In order to estimate the information that ES has gathered about the minimum location, the distribution $\qmin$ is compared to the uniform distribution $u$ by means of the Kullback-Leibler divergence (relative entropy) 
\begin{equation}
\Hp{\qmin} = D_\text{KL}(\qmin || u).
\label{eq:KLdiv}
\end{equation}
The intuition behind this choice is that the uniform distribution contains no information about the location of the minimum ($\Hp{\qmin}=0$), while a very peaked distribution in one location yields large positive values for $\Hp{\qmin}$.  Therefore, $\Hp{\qmin} \ge 0$ is used as a measure of how much we know about the minimum location.

We are not interested in $\Hp{\qmin}$ per se, but in how much this information would increase if we did an experiment on a new location $\uoparam$. Because we cannot know the exact outcome of this experiment at $\uoparam$ beforehand, we have to commit to its expected value, which is feasible using the GP model (see \cref{ssec:GPs}). We define the \emph{expected change in entropy} at an unobserved location $\uoparam$ as
\begin{equation}
\acqui(\uoparam) = \mathbb{E}\left[ \Hp{\qminpred}\right] - \Hp{\qmin}
\label{eq:EdH}
\end{equation}
where $\qminpred$ depends implicitly on $\uoparam$ and represents a new distribution about the minimum if we had evaluated at location $\uoparam$. 
Since the distribution over the minimum $\qminpred$ adds a hypothetical new point to the set of collected observations, it contains a larger amount of information than $\qmin$, which implies $\acqui(\uoparam) \geq 0$. A more detailed definition of~\cref{eq:EdH} can be found in \cite[Sec. 2.6]{HeSc12}. The acquisition function in~\cref{fig:BO} corresponds to an intermediate stage of ES, where
the next point is selected by maximizing $\acqui$ as detailed in \cref{eq:acqui_max}.
\subsubsection{Global minimum computation}
After collecting a new data point, the estimate for the global minimum is found at the minimum of the GP posterior mean. This can be done using local methods with sufficient random restarts, as the GP gradients can be analytically computed.

 \section{Throttle Valve Control}
\label{sec::throttle_valve_control}
Throttle plate controllers are of importance in the automotive industry, where they are used to control the amount of air flowing into the engine's intake manifold. This is especially true for gasoline engines, as the fresh air in the cylinders has to be carefully balanced against the amount of fuel injected into the system to minimize harmful exhaust gas components. Therefore, the need to tune a throttle plate controller is a recurring theme, which has to be revisited anytime a new throttle plate model is used.

\subsection{System description}
The throttle plate consists of a mechanical and an electric part. The electric part includes a DC motor, connected to the throttle plate itself via a transmission. The mechanical part consists of an actual plate attached to two springs with different stiffness and rest positions, resulting in a switching nonlinearity $T_s$ and a second-order nonlinear differential equation of the form
\begin{align}
\dot x_1 &= x_2 \nonumber \\
\dot{x}_2 &= T_s(x_1)+c  x_2 + T_f(x) + b u + d \label{eq:throttle_plate}
\end{align}
where $x_1$ is the opening angle, $c$ a throttle plate parameter, $T_f$ a term representing friction, $u$ the voltage input for the DC motor, and $d$ an external disturbance. 

While it is possible to give an (approximate) analytic expression of $T_s$, this is not necessary for the proposed auto-tuning approach. The nonlinear friction term $T_f$ is usually difficult to model accurately. The term $T_s$ induces a switching behavior due to different spring stiffnesses and rest positions. The switching behavior and the nonlinear friction pose two main challenges of throttle plate control. 

\subsection{Control objectives}\label{subsec::conobj}
The main control objectives for throttle plates are speed and minimal overshoot. These goals were also considered in the indirect RL framework in \cite{BiHgKoMaKn13}.

\textit{Speed} is important in automotive applications because the throttle plate controls the air flow into the engines intake manifold and, for gasoline engines, the amount of fuel injected into the system must usually be proportional to the amount of fresh air. Thus, the amount of fresh air has a critical impact on the (maximal) engine power. A fast throttle plate control is therefore an essential prerequisite for a quickly responding vehicle. The measure for speed used in our experiments is the $T_{90}$-time of the system; that is, the average time that the closed-loop system needs to decrease the absolute control error by 90\% (on a step input).

\textit{Minimal overshoot} is critical near the systems boundaries for safety reasons and, for the rest of the systems domain, it is desirable for reasons of efficiency and/or comfort.

Secondary control objectives include robustness, low noise amplification, and disturbance rejection. 

\subsection{Experimental Setup}
\begin{figure}[tb]
	\centering
		\includegraphics[width=\columnwidth]{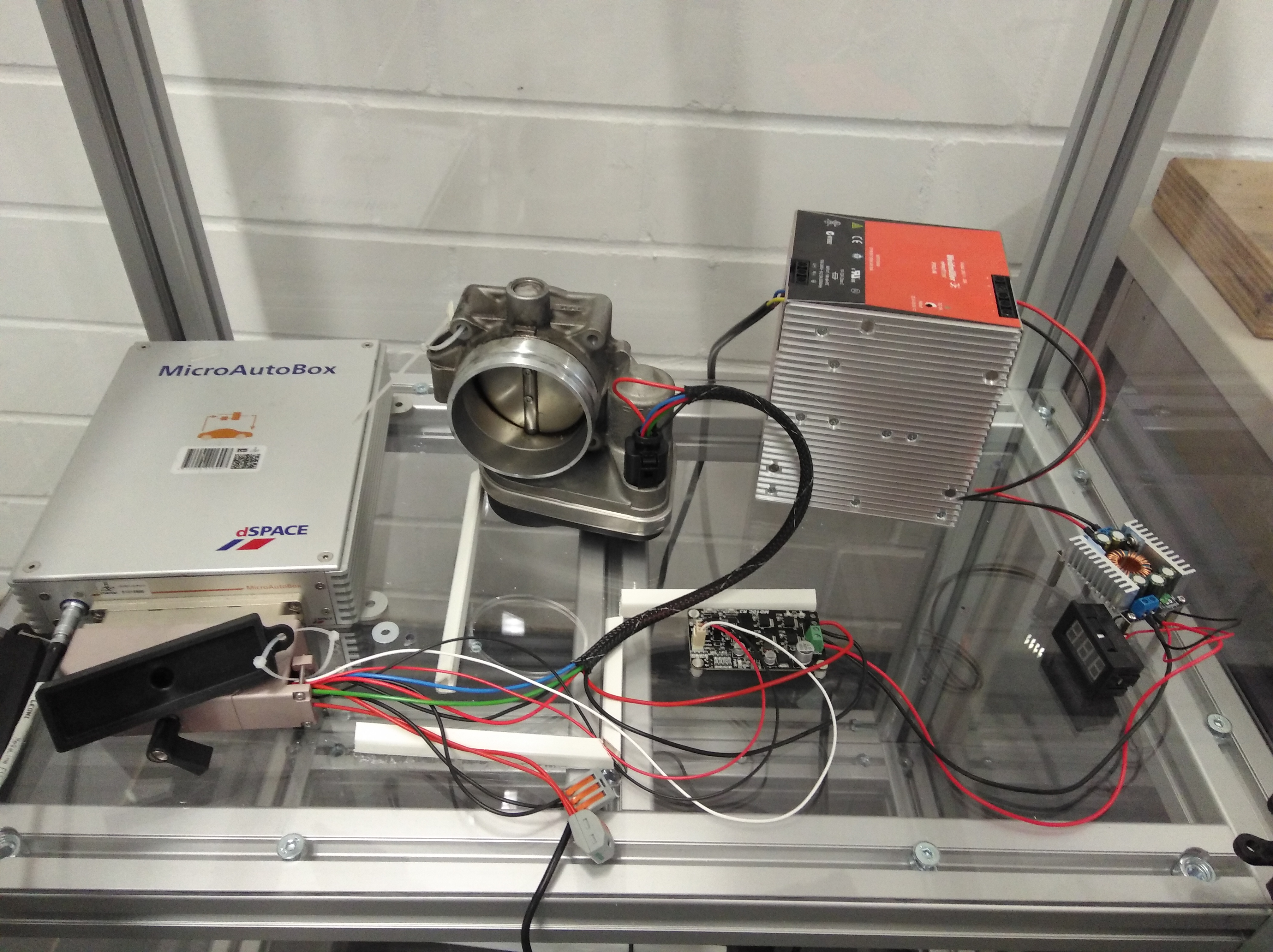}
	\caption{Experimental setup. The trottle plate is shown in the center. It is controlled via the MicroAutoBox (left).}
	\label{fig:experimental_setup}
\end{figure}

The experimental setup used in our experiments consists of a throttle plate, a MicroAutoBox (DS1401), a half-bridge for the DC gain, and required periphery such as power supply; see \cref{fig:experimental_setup}. Communication between Matlab and the DS1401 is implemented using the dSPACE ASAM XIL API. The sampling frequency is set to 1kHz.

 \section{Controller Structure}
\label{sec:controlStructureGeneral}
In this section, we describe the controller used in the throttle valve application, which is an ADRC (active disturbance rejection control) controller, and discuss the reasons for choosing this structure.
We will not discuss the general theory of ADRC since there are already several papers with that aim (i.e.\ \cite{HUANG2014963}; also see \cite{7525139} for the discussion of ADRC of a two-mass spring).
 
The key idea of ADRC is to combine all unknown parts of the system \eqref{eq:throttle_plate} into the \emph{total disturbance}
\begin{equation}
\psi = T_s(x_1) + c x_2 + T_f(x) + d- a^\transp x
\end{equation}
where $a \in \mathbb R^2$, 
and then design an observer-based feedback controller for the \emph{extended system}
\begin{align}
\dot{x}_1 &= x_2 \nonumber \\
\dot{x}_2 &= a^\transp x + \psi + b u \label{eq:sys_state_ext}  \\
\dot{\psi} &= 0 \nonumber\\
y &= x_1. \nonumber
\end{align}
The observer is hence used to estimate the state $x$ and the total disturbance $\psi$.

In theory, the term $a^\transp x$ in Equation \cref{eq:sys_state_ext} can be set to zero by choosing $a = 0$. This would result in a chain of integrators acting on the control signal $u$ and the estimated disturbance $\hat \psi$. 
Since the system \eqref{eq:sys_state_ext} is used for the construction of the observer and the controller, using $a \neq 0$ enables us to use a linear approximation of the system, resulting in better observer and controller performance (at the cost of additional optimization variables). In our experiments, we have found the inclusion of $a$ into \cref{eq:sys_state_ext} critical for performance. 

The nominal system (\ie \cref{eq:sys_state_ext} with $\psi \equiv 0$) is not necessarily a good approximation of the full system decribed in \cref{eq:throttle_plate}, even if the BO optimization produced good results. Rather, the optimization will choose $a$ to support the desired control goals, in the form of the cost functional $\cost$. Thus, ``errors'' in the nominal model are acceptable, or even beneficial, if they help to minimize $\cost$. 

\subsection{Extended state-observer design}
In the physical experiments, we used a linear Luenberger observer constructed via pole placement for the extended system \cref{eq:sys_state_ext}. 
The observer computes estimates $\hat{x}$ and $\hat{\psi}$ of the physical state $x$ and total disturbance $\psi$ in \cref{eq:sys_state_ext}.

Pole placement ensures that the observer is stable and has preassigned nominal dynamics. For practical reasons, we used one (real) optimization variable for all observer poles $\obspoles$, thus all observers have a triple pole at the same stable location.  Other choices are clearly conceivable and do not affect the proposed auto-tuning framework except for possibly increasing the number of parameters.

It is possible to allow for a more general observer structure (\eg a nonlinear one), and such a structure might be beneficial for performance. Since the performance of the test system was satisfactory with the assumptions we made, we did not pursue this direction further.

\subsection{State-feedback controller for nominal system}
We design the control matrix $\ctrmatrix$ for the nominal system (\cref{eq:sys_state_ext} with $\psi \equiv 0$)
via pole placement, in an analogous way as for the observer. This adds the closed-loop poles $\ctrpoles$ to the list of optimization variables. We also add a constant pre-amplification $v$ (calculated to yield a DC gain of one from reference to output, not optimized by BO), yielding the equation $u = \ctrmatrix \hat x+v r$, where $\hat x$ is the estimated system state.

As for the observer, we set all desired poles $\mathcal{P}_\text{ctr}$ to the same value in the physical experiment. Thus, the discussion regarding more complex structures equally applies here.

\subsection{The full controller}
The full ADRC controller is given by the extended state observer and the following state-feedback law:
\begin{equation}
u = \text{sat}\left( \ctrmatrix \hat x + v r - \frac1{b}\hat{\psi} \right)
\label{eq:full_controller}
\end{equation}
where $\text{sat}: \mathbb R \rightarrow \mathbb R, \, x \mapsto \text{sign}(x) \min(|x|,1)$ is a saturation term, introduced to capture the boundedness of the system input.
For the throttle valve controller, we thus have the following tuning parameters: $\theta = (a, \mathcal{P}_\text{obs}, \mathcal{P}_\text{ctr})$.

\subsection{ADRC and throttle plate control}

The decision for ADRC as throttle plate controller over, for example, more typical PID controller structures was made independently of this work.  One main reason is to lessen the effort needed for manual calibration in practice, as is pointed out below.  BO tuning can be applied to other controller structures likewise (see related work in \sect \ref{sec:intro}).

The ADRC control structure is straightforward to tune for humans because of the clear cause-effect relationship of the parameters to the closed-loop behavior: The pole location of the observer determines the speed at which the nonlinear part is estimated and compensated; a too fast pole introduces high frequency modeling errors (due to $\psi$ not being constant) and suffers from sensor noise amplification, whereas a too slow pole will leave the nonlinearities uncompensated at mid to high frequencies. The pole location of the controller determines the overall speed of the closed loop (with obvious limitations from noise and input saturation).
Such an intuitive understanding of cause-effect relationships of the parametrization is critical for manual tuning. It is well known \cite{AsHaHaHo93} that even the tuning of PID controllers is more often than not beyond human capabilities due to its parametrization's non-convex nature. Tuning higher order controllers with unclear cause-effect relationships of the parameters to the closed loop (e.g., tuning the coefficients of the numerator and denominator polynomials of a linear controller) is completely beyond human capabilities when one is left to trial and error.

As with any control method, stability in the presence of sufficiently high uncertanties cannot be guaranteed. In this case, $\psi$ may be considered an uncertainty of a certain (nonlinear) dynamics, wheras the observer assumes this to be a constant value. 
While stability is guaranteed under this assumption ($\psi \equiv \text{const}$), and there is some robustness to the violation of this assumption, a general stability proof for arbitrary uncertainties cannot be given.

In essence, this means that ADRC is suitable for systems whose dynamics are close enough to being integrator chains (which is the case for a throttle plate). To summarize, the reasons for choosing this controller structure are:
\begin{itemize}
	\item This controller structure is well-suited for manual tuning. 
	\item We can use a simple nominal model (in our case: linear, second order, no zeros) to tune the nominal controller and observer, as the nonlinearities will be handled by the extended state. This of course depends critically on the actual system. As the experiments show, the control structure works well in this instance, and one might expect it to generalize to ``similar'' systems.
	\item Using pole placement, this control structure leaves few parameters to tune, which is beneficial for the performance of BO algorithms.
	\item Since we are using a static control matrix $\ctrmatrix$, the controller itself does not have any internal states. Therefore, and because the observer uses the saturated input, windup causes no problems for this control structure.
\end{itemize}

 \section{Design of Learning Experiments} \label{sec::design_learning_experiments}
In this section, we discuss the remaining design choices for applying the proposed tuning algorithm, such as choice of hyperparameters (including kernel functions) and cost functionals. 

The number of iterations for each BO-tuning experiment was set to ten; that is, BO was given a budget of ten experimental evaluations to find the best controller.  All measured data was filtered with a sixth order, non-causal, zero-phase Butterworth filter.

\subsection{Choice of functionals}\label{subsec::choiceofffunctionals}
In our experiments, we compared two functionals, $\costheur$ and $\costprin$, a heuristic and system norm functional. Each of these was tested with both ES (\cref{subsec::ES}) and the Matlab implementation of BO with EI aquisition function (\cref{subsec::EI}).

\subsubsection{The heuristic functional}
This functional was constructed directly from the goal of speed with minimal overshoot. The overshoots and $T_{\text{90}}$-times are calculated via the (controlled) systems response to a reference signal consisting of a series of steps. Each step was held for 2 seconds, and each series of steps lasted 2 minutes. The measured overshoot on step $i$ is denoted by $h_i$, and the measured $T_{\text{90}}$-time on step $i$ by $T_{\text{90},i}$. The \emph{heuristic functional} is defined by
\begin{equation}
\costheur = \frac1n \sum_{i=1}^n{(h_i+T_{\text{90},i})}. \label{eq:J_heur}
\end{equation}

The $T_\text{90}$-time corresponds to system speed (a design goal), and may seem redundant, as we have a parameter $\ctrpoles$ corresponding to the desired \textit{nominal} system speed. However, we did not want to make the assumption that these two values are equivalent because the actual system is quite different from the nominal one.

\subsubsection{The system norm functional}
The idea behind this functional is to use standard system norms, as well as a speed term to allow for a tradeoff between robustness, speed, and overshoot.  For a detailed discussion of system norms, we refer the reader to standard control theory literature, e.g.\ \cite{Skogestad:2005:MFC:1121635}.

The reference for each experiment was a chirp signal with frequencies ranging between 0.1 Hz and 30 Hz and an amplitude of $20^{\circ}$ centered around $25^{\circ}$. Under these circumstances, the system behaves approximately like a linear system, and thus several well-known system norms can be defined.

\begin{figure}[bt] 
	\centering
		\includegraphics[width=\columnwidth]{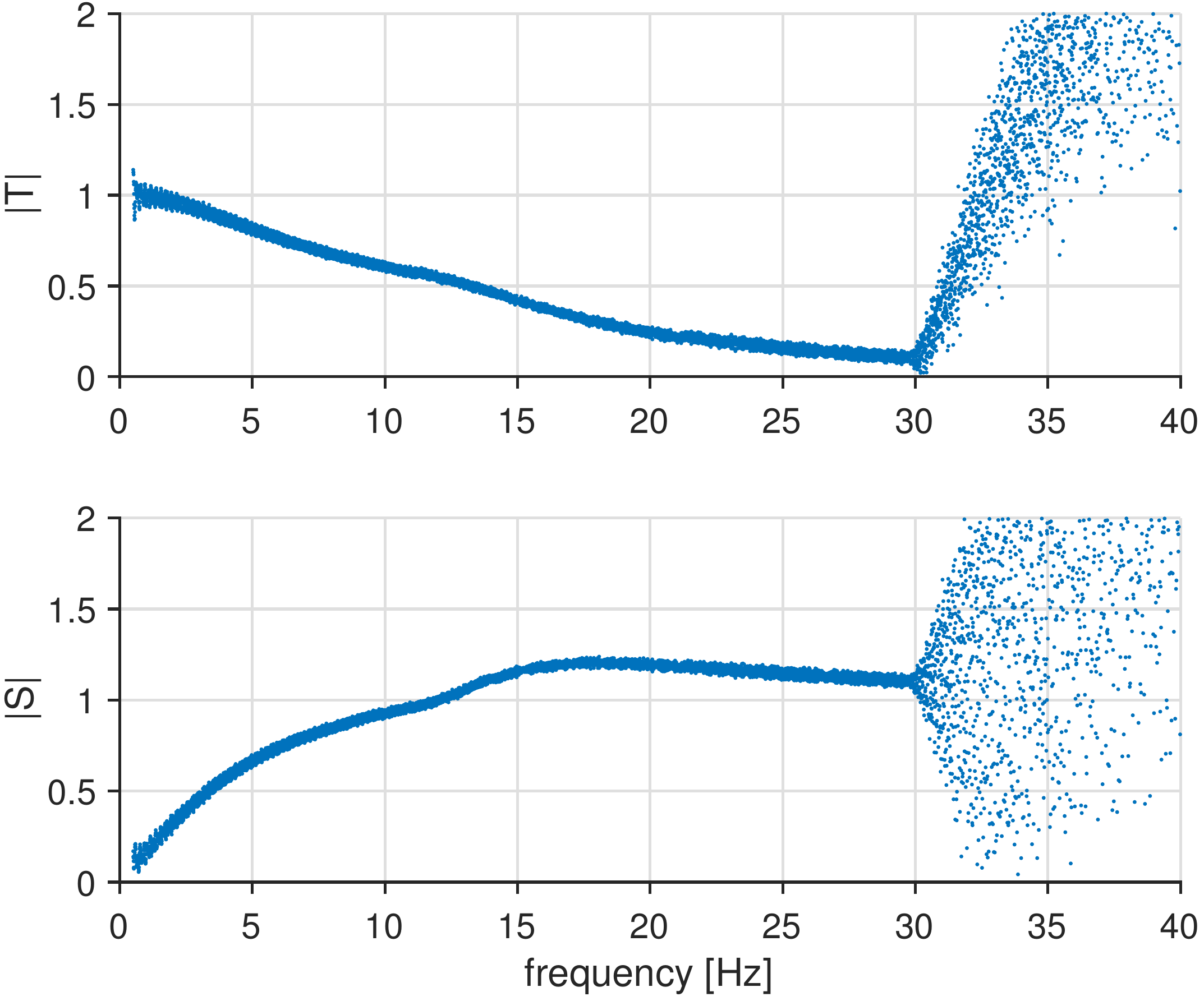}
	\caption{Estimated absolute transfer and sensitivity function.  Only values with frequency in the interval $[0.5\text{Hz},\, 28 \text{Hz}]$ were used in the experiments.}
	\label{fig:functions}
\end{figure}

The throttle plate position signal and reference are used to calculate an approximation of the systems sensitivity $S$ and closed-loop transfer function $T$ (cf. \cref{fig:functions}). In the figure, one can clearly see that the system was not excited beyond 30 Hz. It is important to stress that we do not fit a system here -- we merely apply Fast Fourier Transformation (FFT).

For the system norm functional, we used the $H_\infty$ norm of the senstivity $||S||_\infty$ to provide a measure of robustness, the $H_2$-norm of the transfer function $||T||_2$ to bound the overshoot, and the frequency $f_\text{s}$ where $S$ first crosses the gain $\frac12$, roughly corresponding to a $-3 \, \text{dB}$ gain. The term $e^{\frac{-f_\text{s}}2}$ is a speed term that decreases very slowly for large values of $f_\text{s}$. The system norm functional is thus defined as
\begin{equation}
\costprin = \frac12(||S||_\infty + ||T||_2) + e^{\frac{-f_\text{s}}2}.
\label{eq:J_prin}
\end{equation}

\subsection{Hyperparameters}
The GP prior used in the BO framework depends on
a set of hyperparameters, as explained in \cref{ssec:GPs}. 
Although it is possible to estimate them from data during the experiments, we decided to fix them for the ES method (\cref{subsec::ES}).
Matlab's implementation of EI (\cref{subsec::EI}) always employs the ardmatern52-Kernel (see \cite{Rasmussen2006Gaussian} for a definition of kernels) and its built-in hyperparameter optimization.

The hyperparameters
for ES were generated via maximum likelihood on a set of ten prior measurements. Out of a small set of standard kernels, the ardSE and the ardRQ-Kernel led to the best results in the case of the heuristic and system norm functional, respectively. Therefore, the results discussed in \sect \ref{sec::experimental_results} correspond to that choice. Even though the kernels that produced the best results in our experiment were differentiable, the general approach does not depend on the differentiability of the cost functional $\cost$. Thus, non-differentiable kernels like a $\gamma$-exponential kernel (for $\gamma < 2$) are equally applicable.

Table \ref{tab:length_scales} shows the hyperparameters used for the experiments. Instead of the parameters $a = (a_1, a_2)$ in \cref{eq:sys_state_ext}, we use two (real, stable) poles $P_1$ and $P_2$, and instead of $\obspoles$ and $\ctrpoles$, we use the $T_\text{exp(-6)}$-times $T_\text{obs}$ and $T_\text{set}$, respectively. For any variable $x$, we denote the lengthscale hyperparameter corresponding to $x$ by $l_x$.
We note that, according to the preliminary measurements, the lengthscale $l_{P_1}$ for the system norm functional is so large that the value of $P_1$ is not expected to have a significant effect on the (GP approximation to) the functional $\costprin$. 

\begin{table}[tb]
	\centering
	    \caption{Hyperparameters for different functionals for the Entropy Search algorithm}
		\begin{tabular}{c|ccc}
				\hline
				Hyperparameter & Heuristic (ardSE) & Structured (ardRQ)  \\	
				\hline 
						$l_{T_\text{set}}/ms$ &  77 & 173 \\
						$l_{T_\text{obs}}/ms$  & 13 & 51	\\
						$l_{P_1}$ & 12.3 & $1.07\cdot10^5$ \\
						$l_{P_2}$ & 56.7 & 134\\
						noise std & $1.00\cdot10^{-3}$ & $3.94\cdot10^{-3}$\\
						prior process std & 0.084 & 0.244 \\
						$\alpha$ & - & 0.315
		\end{tabular}
		\label{tab:length_scales}
\end{table}

\subsection{Safety bounds}
The following safety bounds on the parameters $\param$ ensure that parameter combinations cannot harm the hardware:

\begin{description}
	\item[$T_\mathrm{set}$:] Constrained in the interval $[60 \, \text{ms}, 200 \, \text{ms}]$.	Below $60 \, \text{ms}$, the noise would start to pose an (audible) problem, and as speed was a design goal, $200 \, \text{ms}$ was set as an upper bound. 
	\item[$T_\mathrm{obs}$:] Constrained in $[10 \, \text{ms}, 40 \, \text{ms}]$.
	Since it is desireable that the observer speed is somewhat greater than the nominal system speed, we set the upper bound to $40 \, \text{ms}$. The lower bound was set to $10 \, \text{ms}$ because, for lower values, the observer would introduce too much noise into the system.
	\item[$P_1, P_2$:] From insight into the physical process, we expect two stable, real-valued poles in the ranges $[-e^{2},-e^{-1}]$ and $[-e^{5},-e^{2}]$; that is, one slow and one fast pole. 
\end{description}

Since this is effectively a box constraint, and the controller parameters $\param$ are the optimization variables, these constraints provide no additional difficulty for the chosen optimizers.
 \section{Experimental Results}\label{sec::experimental_results}
This section reports the experimental results of applying BO auto-tuning for the throttle valve controller.  In addition to discussing achieved performance with respect to the main objectives \eqref{eq:J_heur} and \eqref{eq:J_prin}, we also consider the secondary objectives (see \cref{subsec::conobj}). 

As mentioned in the introduction, these results are meant to empirically strengthen the claim that BO is an adequate solution to the proposed tuning problem. Our goal is not to systematically compare the different instances of BO (ES and EI) against each other, but to contrast these with manual tuning.

Even though we only list one set of results for each experiment, we have conducted some of the tuning experiments several times.  We did not find any problems with repeatability due to the stochasticity of the optimization. On the other hand, there was a notable dependency on the correct choice of hyperparameters (including the kernels) as discussed in \cref{ssec:BO}.

\subsection{Controllers tuned with the heuristic functional}
Table \ref{tab::heur_main_obj} shows the achieved values of the main control objectives for three different algorithms: manual tuning by a domain expert (including system identification), ES (\cref{subsec::ES}) and (the Matlab implementation of) EI (\cref{subsec::EI}), for the heuristic functional $\costheur$. 

\begin{table}[bt]
	\centering
\caption{Primary control objectives for the heuristic controllers}
		\begin{tabular}{c|c|c|c}
\label{tab::heur_main_obj}
			Algorithm & $\costheur$ & mean $T_{90}$ $[\text{ms}]$ & mean overshoot $[^\circ]$\\
			\hline \hline
			manual tuning & 0.174 & 67 & 0.107\\
			Entropy Search & 0.141 & 53 & 0.088\\
			Expected Improvement & 0.144 & 56 & 0.088 
		\end{tabular}
\end{table}

\begin{table}[bt]
	\centering
\caption{Secondary control objectives for the heuristic controllers}
		\begin{tabular}{c|c|c|c|c}
\label{tab::heur_sec_obj}
			Algorithm & robustness & noise & $T_\text{dist} \, [\text{ms}]$ & $h_\text{dist} \, [^\circ]$ \\
			\hline \hline
			manual tuning & 0.83 & 0.082 & 122 & 3.87 \\
			Entropy Search & 0.83 & 0.078 &77 & 2.17 \\
			Expected Improvement & 0.83 & 0.081 & 90& 2.48
		\end{tabular}
\end{table}

\begin{figure}
	\centering	
		\includegraphics[width=\columnwidth]{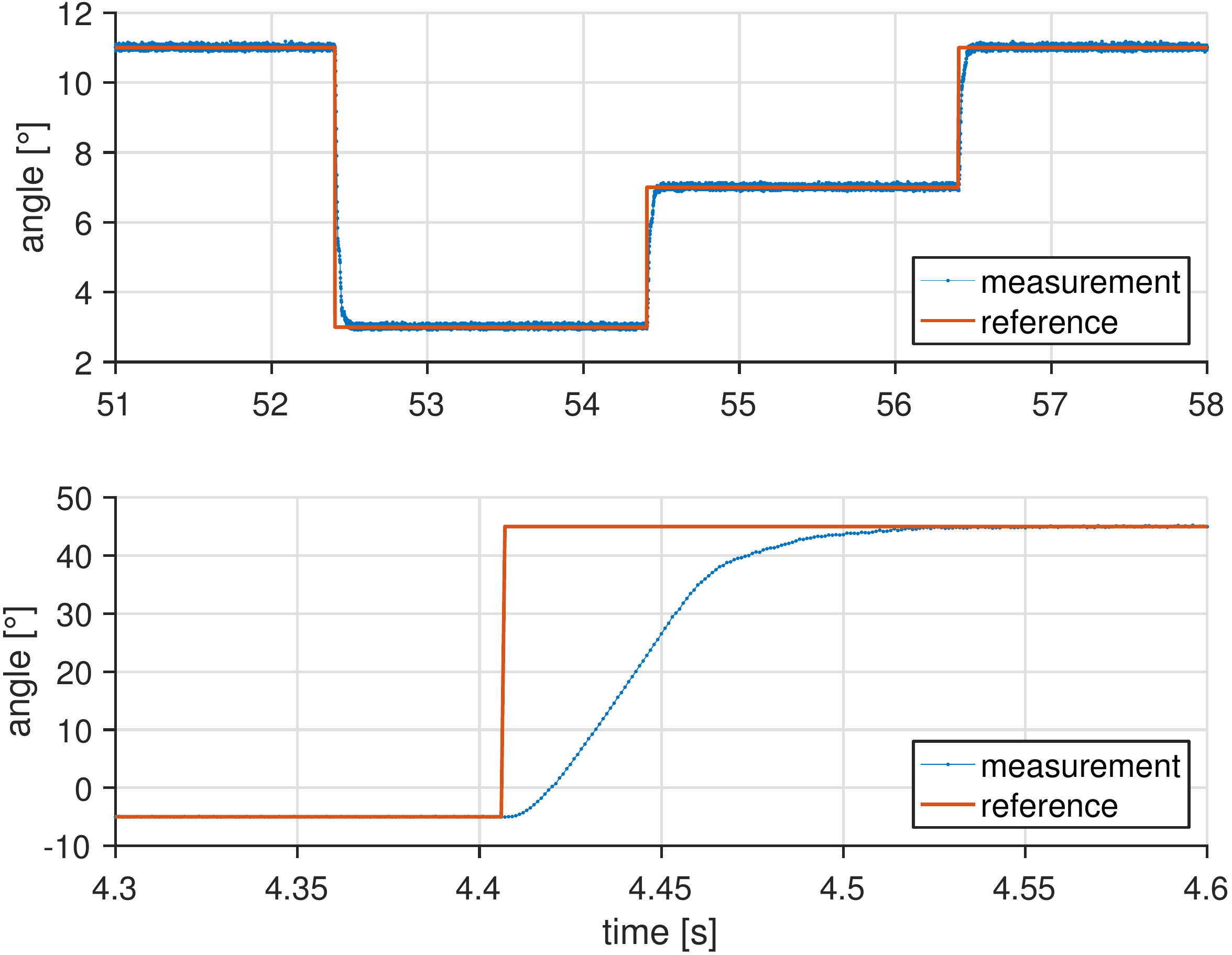}
\caption{Steps measurements for the controller tuned via Entropy Search with the heuristic functional. The lower subfigure shows an earlier step from the same measurement (zoomed-in in time and zoomed-out in angle).}
\label{fig:heuristic_steps}
\end{figure}

A typical series of small steps and one large step for the controller tuned with ES is shown in Figure \ref{fig:heuristic_steps} -- for this plot, the measured values were not filtered. 

The results for the two controllers tuned via BO are comparable. They outperform the nominal controller in both main control objectives. Corresponding to the estimated noise levels for the functionals, the difference between manual tuning and BO is statistically significant.

We chose the following measures for the secondary control objectives (see \cref{subsec::conobj}):
\begin{itemize}
	\item \emph{Robustness} is measured as the inverse of the $H_\infty$-norm of the sensitivity (calculated as described in \sect \ref{sec::design_learning_experiments}). For this application, we consider any value above 0.6 sufficient.
	\item \emph{Noise} is measured  as the mean of the standard deviation of the output of the controlled system at set points $0^\circ, 10^\circ, ...\,, 60^\circ$. Due to the nonlinearity of the system, it is impossible to give just one number that describes noise amplification, which necessitates measurements at different set points.
	\item $T_\text{dist}$ is the duration of disturbance rejection, for a controlled environment set to an angle of $30^\circ$. The disturbance took the form of a step of height 0.7.
	\item $h_\text{dist}$ is the height of the overshoot that the disturbance produced.
\end{itemize}

The obtained values of the secondary objectives are shown in \cref{tab::heur_sec_obj}. They indicate that all controllers have a sufficient robustness measure. The noise standard deviation is comparable over the different controllers, and disturbance rejection is acceptable in all cases. 

The results with BO tuning are better than what is obtained with manual tuning by an expert using system identification and validation measurements. The obtained controller parameters are thus suitable for an industrial application (albeit additional aspects such as safety would also need to be investigated before an actual application, which is not the focus herein). 

\subsection{Controllers tuned with the system norm functional}
Tables \ref{tab::str_main_obj} and \ref{tab::str_sec_obj} show the corresponding values for the controllers tuned with the system norm functional $\costprin$. The numbers indicate that both optimization algorithms have traded robustness for speed. Nonetheless, the robustness measure is kept above the threshold of 0.6. 

While the $H_2$-norm can be used to bound the overshoot (assuming a finite settling time), this bound is generally not tight. Thus, in general, using the $H_2$-norm (as in the system norm functional) will not have the same effect as penalizing the overshoot directly (as in the heuristic functional). Consequently, the different tuning experiments ($\costprin$ vs. $\costheur$) are not directly comparable.

\begin{table}[tb]
	\centering
\caption{Main control objectives for the system norm controllers}
		\begin{tabular}{c|c|c|c}
\label{tab::str_main_obj}
			Algorithm & $\costprin$ & mean $T_{90}$ $[\text{ms}]$ & mean overshoot $[^\circ]$\\
			\hline \hline
			manual tuning & 1.498 & 67 & 0.107\\
			Entropy Search & 1.402 & 46 & 0.318\\
			Expected Improvement & 1.417 & 53 & 0.082 
		\end{tabular}
\end{table}

\begin{table}[tb]
	\centering
\caption{Secondary control objectives for the system norm controllers}
		\begin{tabular}{c|c|c|c|c}
\label{tab::str_sec_obj}
			Algorithm & robustness & noise & $T_\text{dist} \, [\text{ms}]$ & $h_\text{dist} \, [^\circ]$ \\
			\hline \hline
			manual tuning & 0.83 & 0.082 & 122 & 3.87 \\
			Entropy Search & 0.76 & 0.077 & 155 & 4.77 \\
			Expected Improvement & 0.81 & 0.077 & 49 & 1.71
		\end{tabular}
\end{table}

These experiments have shown that it is also possible to define and optimize controllers based on a more system-theoretic viewpoint using system norms like $H_2$ and $H_\infty$ in combination with BO. In particular, this shows that the approach to parameter tuning presented in this paper does not necessarily lead to a ``black box'' system since useful control-theoretic characteristics (such as robustness) can be incorporated into the learning algorithm.

\subsection{Discussion}
The presented experiments show that the proposed tuning method yields competitive results compared to manual tuning.
Apart from the quality of the results as measured by the cost functional, we consider the following aspects relevant advantages of the proposed BO auto-tuning.

\subsubsection*{Structured approach}
The proposed method follows the same steps regardless of the system being tuned, or the person tuning it. Furthermore, with BO, it is clear why and where a measurement was taken (and can be documented accordingly). With manual tuning, the steps taken and the parameters evaluated depend on the engineer, and the choices might not be easily understood and retraced in hindsight.

\subsubsection*{Tuning times} The tuning time of the reported experiments was less than one hour for $\costheur$ and roughly one hour for $\costprin$, including validation measurements. 
In the case of ES, this included ten measurements for hyperparameter optimization. 
In the case of the Matlab implementation of EI, the hyperparameters were estimated online, which further decreased the tuning times. 
In comparison, the time required for manual tuning is highly subjective, but for the specific case herein, the engineer spent significantly 
more time on identification measurements
than the BO algorithm spent on measurements during the complete tuning process including validation measurements.  The manual tuning of the ADRC required, in particular, system identification (for the model used in the observer), selection of appropriate excitation signals, and tuning of remaining controller and observer parameters.
For BO, no explicit system identification step is required and no specific excitation signals need to be chosen.  It can take some time to find an appropriate cost functional, however.  BO parameter learning will involve a similar workflow also for other controller structures; thus, we expect it to be generally advantageous over manual tuning in terms of time.

\subsubsection*{Generalization} We expect the BO tuning process to generalize well to similar systems, for example, when controllers for several throttle valves are to be tuned in succession.  This will yield significant time savings as the BO tuning can be executed with no or minimal changes for different throttle valves.

\subsubsection*{Different functionals}  We have empirically shown that the BO tuning process can handle widely different functionals and thus control objectives. This is relevant in industrial practice, for example, if some requirements change after an initial tuning. The same setup with an adapted functional can be applied, and the measurements taken before the change are able to serve as prior information for the GP model of the cost function (and for hyperparameter estimation). 

\subsubsection*{Transferability}  In general, manual tuning requires some intuition about the dynamic system and the parameters. In contrast, the BO process is easily transferable between engineers because the assumptions are explicitly programmed into the BO code (via the choice of functional, hyperparameters, kernels, etc.).

 \section{Concluding Remarks}
Direct automatic controller tuning via Bayesian optimization (BO) is proposed herein.  The developed framework is applicable to diverse controller tuning problems (\eg different controller structures, and objectives) and has been shown to be highly data-efficient; competitive throttle valve controllers were obtained from only about ten trials.  

Fundamentally, in data-based optimization or tuning, no guarantee for (global) optimality can be given without prior assumptions.  If the objective function can potentially vary arbitrarily, one cannot  give any reliable estimate about its shape and thus its minimum.  Gaussian process (GP) models, as employed herein, allow for specifying properties of the objective function via a probabilistic prior.  Under such a model, BO can reason about function minima in a probabilistic sense, which is used here to effectively choose next evaluation points and address global optimization problems.  

In addition to the GP prior, we also leverage elementary process knowledge in the presented approach.
In the ADRC structure, a crude dynamics model, which essentially includes knowledge of the system order only, is combined with an observer to capture unmodeled nonlinear dynamics and disturbances.  The key parameters (dynamics poles, controller and observer gain) are learned from experimental data via BO.  This combination of ADRC with BO thus provides a very flexible structure that strikes a meaningful balance between physical knowledge and learning from data.  
The approach is flexible in that additional process knowledge could be included (\eg partially known dynamics), or omitted (\eg no dynamics poles). We thus expect this approach to be effective for a large class of practical control problems.  

BO is an active area of research and is becoming increasingly popular in diverse application areas (see \cite{shahriari2016taking}).  This article is the first to demonstrate the potential of BO for automatic controller tuning in industry.  
Current and future research 
concerns improving sample-efficiency by exploiting multiple information sources \cite{MaBeHeScKrScTr17}, design of problem-specific kernels \cite{MaHeScTr17}, and extensions to high-dimensional problems.  These all aim at making BO controller tuning an even more powerful practical tool.

\section*{Acknowledgment}
The authors would like to thank Manus Thiel (IAV GmbH) for his help while setting up the experiments.

\bibliographystyle{IEEEtran}
\bibliography{Database}

\begin{thebibliography}{10}
\providecommand{\url}[1]{#1}
\csname url@samestyle\endcsname
\providecommand{\newblock}{\relax}
\providecommand{\bibinfo}[2]{#2}
\providecommand{\BIBentrySTDinterwordspacing}{\spaceskip=0pt\relax}
\providecommand{\BIBentryALTinterwordstretchfactor}{4}
\providecommand{\BIBentryALTinterwordspacing}{\spaceskip=\fontdimen2\font plus
\BIBentryALTinterwordstretchfactor\fontdimen3\font minus
  \fontdimen4\font\relax}
\providecommand{\BIBforeignlanguage}[2]{{%
\expandafter\ifx\csname l@#1\endcsname\relax
\typeout{** WARNING: IEEEtran.bst: No hyphenation pattern has been}%
\typeout{** loaded for the language `#1'. Using the pattern for}%
\typeout{** the default language instead.}%
\else
\language=\csname l@#1\endcsname
\fi
#2}}
\providecommand{\BIBdecl}{\relax}
\BIBdecl

\bibitem{Je06}
M.~Jelali, ``An overview of control performance assessment technology and
  industrial applications,'' \emph{Control Engineering Practice}, vol.~14,
  no.~5, pp. 441--466, 2006.

\bibitem{AsHaHaHo93}
K.~J. {\AA}str{\"o}m, T.~H{\"a}gglund, C.~C. Hang, and W.~K. Ho, ``Automatic
  tuning and adaptation for {PID} controllers-a survey,'' \emph{Control
  Engineering Practice}, vol.~1, no.~4, pp. 699--714, 1993.

\bibitem{Ha15}
T.~H{\"a}gglund, ``Autotuning,'' in \emph{Encyclopedia of Systems and Control},
  J.~Baillieul and T.~Samad, Eds.\hskip 1em plus 0.5em minus 0.4em\relax
  Springer, 2015, pp. 50--55.

\bibitem{Hj02}
H.~Hjalmarsson, ``Iterative feedback tuning---an overview,''
  \emph{International Journal of Adaptive Control and Signal Processing},
  vol.~16, no.~5, pp. 373--395, 2002.

\bibitem{KiKr06}
N.~J. Killingsworth and M.~Krstic, ``{PID} tuning using extremum seeking:
  online, model-free performance optimization,'' \emph{IEEE Control Systems},
  vol.~26, no.~1, pp. 70--79, 2006.

\bibitem{FlPu02}
P.~J. Fleming and R.~C. Purshouse, ``Evolutionary algorithms in control systems
  engineering: a survey,'' \emph{Control engineering practice}, vol.~10,
  no.~11, pp. 1223--1241, 2002.

\bibitem{Rasmussen2006Gaussian}
C.~E. Rasmussen and C.~K. Williams, \emph{\BIBforeignlanguage{English}{Gaussian
  Processes for Machine Learning}}.\hskip 1em plus 0.5em minus 0.4em\relax {MIT
  Press}, 2006.

\bibitem{krige1951statistical}
D.~G. Krige, ``A statistical approach to some basic mine valuation problems on
  the witwatersrand,'' \emph{Journal of the Southern African Institute of
  Mining and Metallurgy}, vol.~52, no.~6, pp. 119--139, 1951.

\bibitem{kushner1964new}
H.~J. Kushner, ``A new method of locating the maximum point of an arbitrary
  multipeak curve in the presence of noise,'' \emph{Journal of Basic
  Engineering}, vol.~86, no.~1, pp. 97--106, 1964.

\bibitem{shahriari2016taking}
B.~Shahriari, K.~Swersky, Z.~Wang, R.~P. Adams, and N.~de~Freitas, ``Taking the
  human out of the loop: A review of {B}ayesian optimization,''
  \emph{Proceedings of the IEEE}, vol. 104, no.~1, pp. 148--175, 2016.

\bibitem{HUANG2014963}
Y.~Huang and W.~Xue, ``Active disturbance rejection control: Methodology and
  theoretical analysis,'' \emph{ISA Transactions}, vol.~53, no.~4, pp. 963 --
  976, 2014, disturbance Estimation and Mitigation.

\bibitem{calandra2016bayesian}
R.~Calandra, A.~Seyfarth, J.~Peters, and M.~P. Deisenroth, ``Bayesian
  optimization for learning gaits under uncertainty,'' \emph{Annals of
  Mathematics and Artificial Intelligence}, vol.~76, no. 1-2, pp. 5--23, 2016.

\bibitem{VoTrMaFiPa18}
A.~von Rohr, S.~Trimpe, A.~Marco, P.~Fischer, and S.~Palagi, ``Gait learning
  for soft microrobots controlled by light fields,'' in \emph{Proc. of the
  IEEE/RSJ Int. Conf. on Intelligent Robots and Systems}, 2018, under review.

\bibitem{ScAt10}
S.~Schaal and C.~Atkeson, ``Learning control in robotics,'' \emph{IEEE Robotics
  Automation Magazine}, vol.~17, no.~2, pp. 20--29, Jun. 2010.

\bibitem{KoBaPe13}
J.~Kober, J.~A. Bagnell, and J.~Peters, ``Reinforcement learning in robotics: A
  survey,'' \emph{The International Journal of Robotics Research}, 2013.

\bibitem{doerr_corl_2017}
A.~Doerr, C.~Daniel, D.~Nguyen-Tuong, A.~Marco, S.~Schaal, M.~Toussaint, and
  S.~Trimpe, ``Optimizing long-term predictions for model-based policy
  search,'' in \emph{Proc. of the 1st Conference on Robot Learning}, Nov. 2017,
  pp. 227--238.

\bibitem{DeRa11}
M.~Deisenroth and C.~E. Rasmussen, ``Pilco: A model-based and data-efficient
  approach to policy search,'' in \emph{Proceedings of the 28th International
  Conference on machine learning (ICML-11)}, 2011, pp. 465--472.

\bibitem{DeFoRa15}
M.~P. Deisenroth, D.~Fox, and C.~E. Rasmussen, ``Gaussian processes for
  data-efficient learning in robotics and control,'' \emph{IEEE Transactions on
  Pattern Analysis and Machine Intelligence}, vol.~37, no.~2, pp. 408--423,
  2015.

\bibitem{DoNgMaScTr17}
A.~Doerr, D.~Nguyen-Tuong, A.~Marco, S.~Schaal, and S.~Trimpe, ``Model-based
  policy search for automatic tuning of multivariate {PID} controllers,'' in
  \emph{IEEE International Conference on Robotics and Automation}, May 2017,
  pp. 5295--5301.

\bibitem{BeScKr16}
F.~Berkenkamp, A.~P. Schoellig, and A.~Krause, ``Safe controller optimization
  for quadrotors with {G}aussian processes,'' in \emph{IEEE International
  Conference on Robotics and Automation}, May 2016, pp. 491--496.

\bibitem{MaHeBoScTr16}
A.~Marco, P.~Hennig, J.~Bohg, S.~Schaal, and S.~Trimpe, ``Automatic {LQR}
  tuning based on {G}aussian process global optimization,'' in \emph{IEEE
  International Conference on Robotics and Automation}, May 2016, pp. 270--277.

\bibitem{CaLeSa02}
M.~C. Campi, A.~Lecchini, and S.~M. Savaresi, ``Virtual reference feedback
  tuning: a direct method for the design of feedback controllers,''
  \emph{Automatica}, vol.~38, no.~8, pp. 1337--1346, 2002.

\bibitem{CaSa06}
M.~C. Campi and S.~M. Savaresi, ``Direct nonlinear control design: The virtual
  reference feedback tuning ({VRFT}) approach,'' \emph{IEEE Transactions on
  Automatic Control}, vol.~51, no.~1, pp. 14--27, 2006.

\bibitem{PaDeJaPe06}
D.~Pavkovi{\'c}, J.~Deur, M.~Jansz, and N.~Peri{\'c}, ``Adaptive control of
  automotive electronic throttle,'' \emph{Control Engineering Practice},
  vol.~14, no.~2, pp. 121--136, 2006.

\bibitem{WaYuWaYa13}
H.~Wang, X.~Yuan, Y.~Wang, and Y.~Yang, ``Harmony search algorithm-based
  fuzzy-pid controller for electronic throttle valve,'' \emph{Neural Computing
  and Applications}, vol.~22, no.~2, pp. 329--336, 2013.

\bibitem{BiHgKoMaKn13}
B.~Bischoff, D.~Nguyen-Tuong, T.~Koller, H.~Markert, and A.~Knoll, ``Learning
  throttle valve control using policy search,'' in \emph{Joint European
  Conference on Machine Learning and Knowledge Discovery in Databases}, 2013,
  pp. 49--64.

\bibitem{HeSc12}
P.~Hennig and C.~J. Schuler, ``Entropy search for information-efficient global
  optimization,'' \emph{The Journal of Machine Learning Research}, vol.~13,
  no.~1, pp. 1809--1837, 2012.

\bibitem{jones1998efficient}
D.~R. Jones, M.~Schonlau, and W.~J. Welch, ``Efficient global optimization of
  expensive black-box functions,'' \emph{Journal of Global optimization},
  vol.~13, no.~4, pp. 455--492, 1998.

\bibitem{snoek2012practical}
J.~Snoek, H.~Larochelle, and R.~P. Adams, ``Practical {B}ayesian optimization
  of machine learning algorithms,'' in \emph{Advances in neural information
  processing systems}, 2012, pp. 2951--2959.

\bibitem{gelbart2014bayesian}
M.~A. Gelbart, J.~Snoek, and R.~P. Adams, ``Bayesian optimization with unknown
  constraints,'' in \emph{Conference on Uncertainty in Artifficial
  Intelligence}, 2014, pp. 250--259.

\bibitem{lizotte2008}
D.~Lizotte, ``Practical {B}ayesian optimization,'' 2008, ph.D. thesis.

\bibitem{srinivas2009gaussian}
N.~Srinivas, A.~Krause, S.~M. Kakade, and M.~Seeger, ``Gaussian process
  optimization in the bandit setting: {N}o regret and experimental design,'' in
  \emph{International Conference on Machine Learning}, 2010.

\bibitem{7525139}
H.~Zhang, S.~Zhao, and Z.~Gao, ``An active disturbance rejection control
  solution for the two-mass-spring benchmark problem,'' in \emph{2016 American
  Control Conference (ACC)}, July 2016, pp. 1566--1571.

\bibitem{Skogestad:2005:MFC:1121635}
S.~Skogestad and I.~Postlethwaite, \emph{Multivariable Feedback Control:
  Analysis and Design}.\hskip 1em plus 0.5em minus 0.4em\relax John Wiley \&
  Sons, 2005.

\bibitem{MaBeHeScKrScTr17}
A.~Marco, F.~Berkenkamp, P.~Hennig, A.~P. Schoellig, A.~Krause, S.~Schaal, and
  S.~Trimpe, ``Virtual vs. real: Trading off simulations and physical
  experiments in reinforcement learning with {B}ayesian optimization,'' in
  \emph{IEEE International Conference on Robotics and Automation}, May 2017,
  pp. 1557--1563.

\bibitem{MaHeScTr17}
A.~Marco, P.~Hennig, S.~Schaal, and S.~Trimpe, ``On the design of {LQR} kernels
  for efficient controller learning,'' in \emph{IEEE Conference on Decision and
  Control}, Melbourne, Australia, 2017, accepted.

\end{thebibliography}

\begin{IEEEbiography}[{\includegraphics[width=1in,height=1.25in,clip,keepaspectratio]{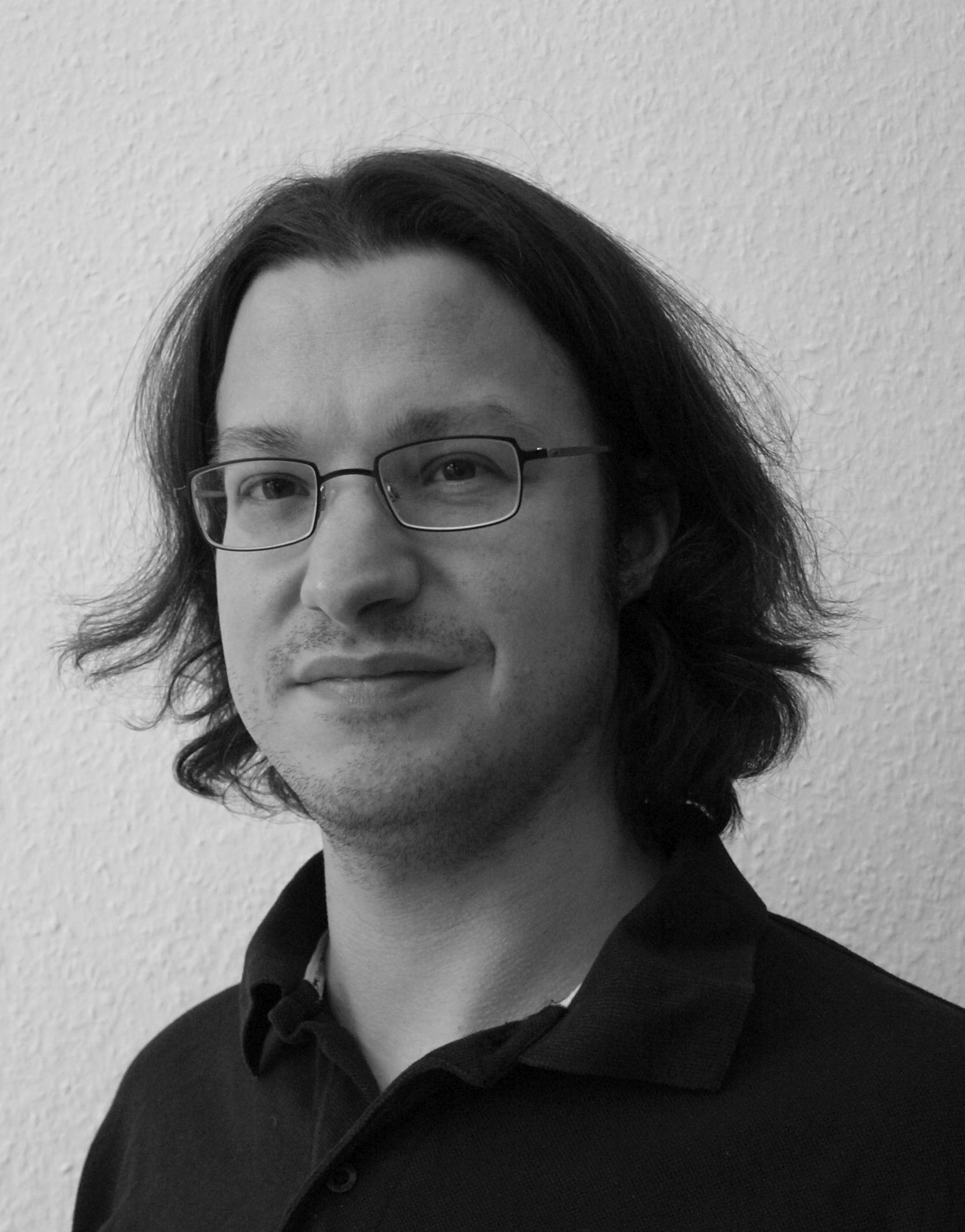}}]{Matthias Neumann-Brosig}
recieved his diploma in mathematics from TU Dortmund in 2010 and his Ph.D. degree (Dr. rer. nat.) in abstract algebra from TU Braunschweig in 2015. Since 2015 he is an employee of IAV GmbH, where he has worked in the Control Engineering Excellence Cluster and the Data Analytics team. 

His research interests are category theory, abstract algebra and applications of pure mathematics in other fields, \eg control theory. 
\end{IEEEbiography}

\begin{IEEEbiography}[{\includegraphics[width=1in,height=1.25in,clip,keepaspectratio]{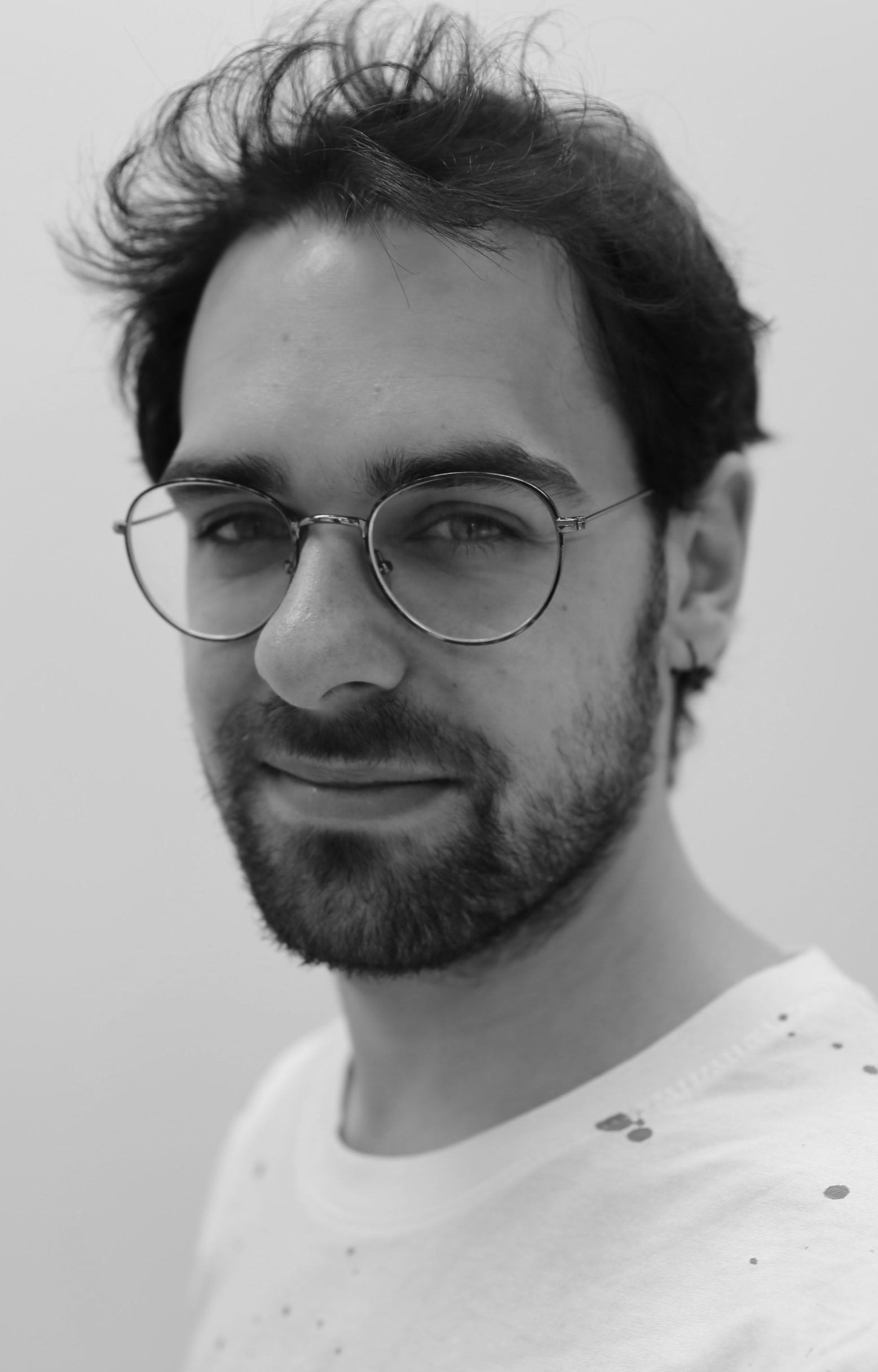}}]{Alonso Marco}
received his M.Sc. degree in automatic control and robotics from Polytechnic University of Catalonia, Barcelona, Spain, in 2015. 

He is currently a Ph.D. student in the Autonomous Motion Department at the Max Planck Institute for Intelligent Systems, in T\"ubingen, Germany. His research interests are in Bayesian optimization, optimal control, and reinforcement learning.

A. Marco obtained a grant for excellent students to study his Master's degree, from Catalunya-La Pedrera Foundation in 2015.
\end{IEEEbiography}

\begin{IEEEbiography}[{\includegraphics[width=1in,height=1.25in,clip,keepaspectratio]{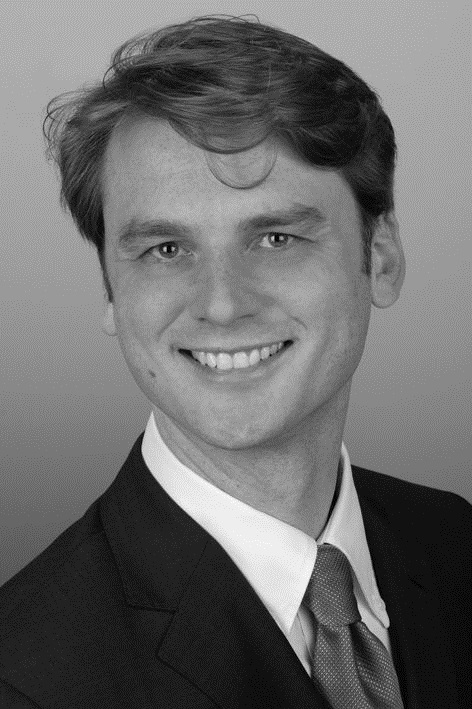}}]{Dieter Schwarzmann}
received the M.Sc. degree in mechanical engineering from Rose-Hulman Institute of Technology and a M.Sc. degree in engineering cybernetics from the University of Stuttgart in 2003. He received his Ph.D. (Dr. sc.) from the Ruhr-University of Bochum in 2007.
He has worked at BOSCH Group from 2006 in various roles, many involving control engineering. In the years 2015 to 2017 he headed the excellence cluster of control engineering at IAV GmbH. Since late 2017 he has taken the role of the central domain expert on software and control engineering at the BOSCH Group. He has been teaching adaptive control at the University of Stuttgart since 2013.
He is focused on control methods for effective application in industry, particularly robust control. His research interests include adaptive control and system identification.
\end{IEEEbiography}

\begin{IEEEbiography}[{\includegraphics[width=1in,height=1.25in,clip,keepaspectratio]{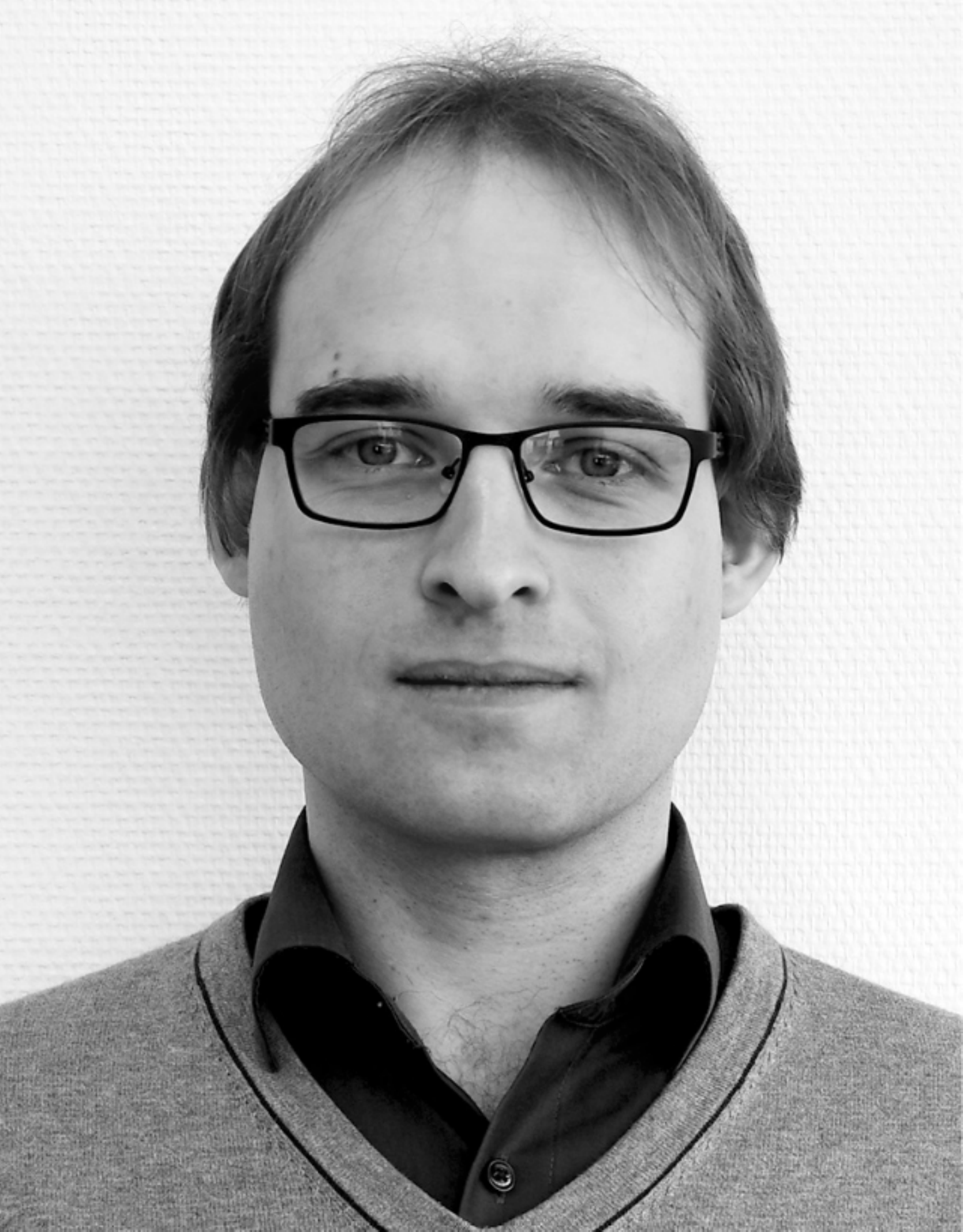}}]{Sebastian Trimpe}
received the B.Sc. degree
in general engineering science and the M.Sc.
degree (Dipl.-Ing.) in electrical engineering from
Hamburg University of Technology, Hamburg,
Germany, in 2005 and 2007, respectively, and the
Ph.D. degree (Dr. sc.) in mechanical engineering
from ETH Zurich, Zurich, Switzerland, in 2013.

He is currently a Research Group Leader at the Max Planck Institute for Intelligent Systems, Stuttgart, Germany, where he leads the independent Max Planck Research Group on Intelligent Control Systems.
His main research interests are in systems and control theory, machine learning, networked and autonomous systems.

Dr. Trimpe is a recipient of the General Engineering Award for the best undergraduate
degree (2005), a scholarship from the German National Academic
Foundation (2002 to 2007), the triennial IFAC World Congress Interactive
Paper Prize (2011), and the Klaus Tschira Award for public understanding of science (2014).
\end{IEEEbiography}

\vfill

\end{document}